\documentclass[amsmath,amssymb,12pt,superscriptaddress,nofootinbib]{revtex4}

\usepackage{graphicx}
\usepackage{bm}

\newcommand{\dd}{\mathrm{d}}

\newcommand{\del}{\partial}

\begin{document}
\baselineskip5.5mm
\thispagestyle{empty}

{\baselineskip0pt
\leftline{\baselineskip14pt\sl\vbox to0pt{
               \hbox{\it Division of Particle and Astrophysical Science, Nagoya University} 
              %\hbox{\it Nagoya University}
               \vspace{1mm}
              \hbox{\it Instituto Superior T\'ecnico}
                             \vss}}
\rightline{\baselineskip16pt\rm\vbox to20pt{
            %\hbox{YITP-1x-xx}
\vss}}%
}

\author{Chul-Moon~Yoo}\email{yoo@gravity.phys.nagoya-u.ac.jp}
\affiliation{
Gravity and Particle Cosmology Group,
Division of Particle and Astrophysical Science,
Graduate School of Science, Nagoya University, 
Nagoya 464-8602, Japan
}

\author{Hirotada~Okawa}\email{hirotada.okawa@ist.utl.pt}
\affiliation{ 
CENTRA, Departamento de F\'isica, Instituto Superior T\'ecnico, 
Avenida Rovisco Pais no 1, 1049-001, Lisboa Portugal
}

\vskip2cm
\title{Black Hole Universe with $\Lambda$}

\begin{abstract}
Time evolution of a black hole lattice universe with a 
positive cosmological constant $\Lambda$ 
is simulated. 
The vacuum Einstein equations 
are numerically solved in 
a cubic box with a black hole in the center. 
Periodic boundary conditions 
on all pairs of opposite faces are imposed. 
Configurations of marginally trapped surfaces are analyzed. 
We describe the time evolution of not only black hole horizons, but also 
cosmological horizons. 
Defining the effective scale factor by using the 
area of a surface of the cubic box, 
we compare it with that in the spatially 
flat dust dominated FLRW universe with the same value of $\Lambda$. 
It is found that the behaviour of the effective scale factor is 
well approximated by that in the FLRW universe. 
Our result suggests that 
local inhomogeneities 
do not significantly affect the 
global expansion law of the universe irrespective of the value of $\Lambda$. 
\end{abstract}


\maketitle
\pagebreak

%%%%%%%%%%%%%%%%%%%%%%%%%%%%%%%%%%%%%%%%%%%%%%%%%%%%%%%%%%%%%%%%
\section{introduction}
\label{sec:intro}
%%%%%%%%%%%%%%%%%%%%%%%%%%%%%%%%%%%%%%%%%%%%%%%%%%%%%%%%%%%%%%%%
%{\bf 
%Abount balck hole lattice universe
%}

The so-called ``black hole lattice universe" has been 
firstly investigated by Lindquist and Wheeler in 1957\cite{RevModPhys.29.432}. 
They regularly arranged $N$ potions of the Schwarzschild spacetime on a 
virtual 3-sphere ($N=5$, 8, 16, 24, 120 and 600), 
and discussed evolution of this lattice universe 
based on the intuitively derived junction conditions 
between the Schwarzschild shell and the 3-sphere. 
The black hole lattice universe is often 
used as one of tools to evaluate effects of local non-linear inhomogeneities on 
the global expansion. 
Recently, black hole lattice universe models 
have been revisited by several authors\cite{Yoo:2013yea,Yoo:2012jz,Clifton:2012qh,Clifton:2009jw,Uzan:2010nw,Bentivegna:2012ei,Bentivegna:2013xna,Bruneton:2012cg,Clifton:2013jpa,Clifton:2014lha,Korzynski:2013tea}. 
Time symmetric initial data for $N$-black hole systems on a virtual 3-sphere 
have been analyzed in Refs.~\cite{Clifton:2012qh,Korzynski:2013tea}. 
Time evolution of the 8-black hole system has been performed 
and analyzed in Ref.~\cite{Bentivegna:2012ei}. 
Initial data for a black hole inside a cubic box with a periodic boundary condition 
have been constructed and analyzed in Refs.~\cite{Yoo:2012jz,Bentivegna:2013xna}, 
and those time evolutions have been investigated in Refs.~\cite{Yoo:2013yea,Bentivegna:2012ei}. 
We call this cubic lattice model the ``black hole universe" in this paper. 
The purpose of this paper is to extend the black hole universe so that it 
admits a positive cosmological constant. 

%{\bf
%Importance of including cosmological constant.
%}

In Ref.~\cite{Yoo:2013yea}, it has been reported that, 
if the box size of the black hole universe 
is sufficiently larger than the horizon radius, 
the global expansion law can be 
well approximated by that in the Einstein-de Sitter universe. 
Our final purpose is to check this fact with a positive cosmological constant. 
Since our universe is likely to be filled with dark energy components, such as 
the positive cosmological constant,  
it is important to investigate the 
effect of local non-linear inhomogeneities on the global expansion law 
with the cosmological constant. 
In the all references listed above, the cosmological constant is 
set to be zero. 
Therefore, solving technical problems to consider non-zero cosmological constant cases, 
we investigate it in this paper.

%{\bf
%Technical significance.
%}

One of non-trivial technical problems is  
how to construct an initial data set 
which is appropriate as a initial condition for 
the time evolution. 
In this paper, we describe a procedure 
to construct puncture initial data for the black hole universe 
with a positive cosmological constant. 
Another interesting problem is to find 
different kinds of marginal surfaces. 
As in the case of Kottler(Schwarzschild-de Sitter) solution, 
the black hole universe with a positive cosmological constant 
can have not only black hole horizons but also 
de Sitter cosmological horizons. 
As far as we know, it is first time 
to numerically find the ${\bf S}^2$ cosmological horizons without any symmetry
which makes it possible to reduce the number of the effective dimension. 
To check the existence and structure of marginal surfaces is very 
useful to understand the spacetime structure. 

%{\bf
%Organization.
%}

This paper is organized as follows. 
In Sec.~\ref{sec:initial}, 
we describe how to construct initial data 
of the black hole universe with a positive cosmological constant. 
Then, we analyze the structure of the initial data 
in Sec.~\ref{sec:marginal} searching for 
different kinds of marginal surfaces. 
In Sec.~\ref{sec:evolve}, 
time evolutions are described. 
The evolution of the configuration of marginal surfaces and the 
expansion law are discussed there. 
Sec.~\ref{sec:summary} is devoted to a summary. 

In this paper, we use the geometrized units in which 
the speed of light and Newton's gravitational constant are one, 
respectively. 

%%%%%%%%%%%%%%%%%%%%%%%%%%%%%%%%%%%%%%%%%%%%%%%%%%%%%%%%%%%%%%%%%
\section{Initial Data }
\label{sec:initial}
%%%%%%%%%%%%%%%%%%%%%%%%%%%%%%%%%%%%%%%%%%%%%%%%%%%%%%%%%%%%%%%%%

\subsection{Constraint equations}

Let us consider solutions of 
vacuum Einstein equations with a positive cosmological constant $\Lambda$ 
described by 
the intrinsic metric $\gamma_{ij}$ and the extrinsic curvature 
$K_{ij}$. 
The Hamiltonian constraint and the momentum constraint equations are
given by 
\begin{eqnarray}
&&\mathcal R+K^2-K_{ij}K^{ij}-2\Lambda=0, 
\label{hamicon1}
\\
&&D_jK^j_{~i}-D_iK=0, 
\label{momcon1}
\end{eqnarray}
where $\mathcal R$ and $D_i$ are the Ricci scalar curvature 
and the covariant derivative with respect to $\gamma_{ij}$, 
and $K=\gamma^{ij}K_{ij}$. 
We perform conformal decomposition in a conventional way as follows:
\begin{eqnarray}
\gamma_{ij}&=&\Psi^4\tilde\gamma_{ij}, \\
K^{ij}&=&\Psi^{-10}\left[\tilde D^iX^j+\tilde D^jX^i-\frac{2}{3}\tilde \gamma^{ij}
\tilde D_k X^k
+\hat A^{ij}_{\rm TT}\right]+\frac{1}{3}\Psi^{-4}\tilde \gamma^{ij}K, 
\label{decomposition}
\end{eqnarray}
where 
$\Psi:=\left(\det \gamma_{ij}\right)^{(1/12)}$, 
$\tilde D_i$ is covariant derivative with 
respect to the conformal metric $\tilde \gamma_{ij}$, and 
$\hat A^{ij}_{\rm TT}$ satisfies 
\begin{equation}
\tilde D_j\hat A^{ij}_{\rm TT}=0~,~~\tilde \gamma_{ij}\hat A_{\rm TT}^{ij} =0. 
\end{equation}

To minimize effects of artificial gravitational radiation, 
we assume
\begin{eqnarray}
\tilde \gamma_{ij}&=&\delta_{ij}, \label{conf-metric}\\
\hat A^{ij}_{\rm TT}&=&0, 
\end{eqnarray}
where $\delta_{ij}$ is the Kronecker's delta. 
Then, from Eqs.~\eqref{hamicon1} and \eqref{momcon1}, 
we obtain 
\begin{eqnarray}
\triangle\Psi&+&\frac{1}{8}(\tilde L X)_{ij}
(\tilde L X)^{ij}\Psi^{-7}-\frac{1}{12}K^2\Psi^5+\frac{1}{4}\Lambda \Psi^5=0, 
\label{eq:hamicon3}
\\
\triangle X^i&+&\frac{1}{3}\partial^i\partial_jX^j-\frac{2}{3}\Psi^6\partial^iK
=0, 
\label{eq:momcon3}
\end{eqnarray}
where 
\begin{equation}
(\tilde L X )^{ij}:=\del ^i X^j+\del ^j X^i-\frac{2}{3}\delta^{ij}\del_k X^k. 
\end{equation}
First, we need to solve these constraint equations 
in appropriate settings for puncture initial data with $\Lambda>0$. 

\subsection{Puncture structure with $\Lambda$}

In this paper, we adopt the Cartesian coordinate system 
$\bm x=(x,y,z)$. 
We consider a cubic region $\mathcal D$ given by 
$-L\leq x \leq L$, $-L\leq y \leq L$ and $-L\leq z \leq L$ 
with periodic boundary conditions 
on all pairs of faces opposite to each other. 
Thus, the domain $\mathcal D$ is homeomorphic to the 3-torus 
${\bf T}^3$. 
The black hole is represented by 
a structure like the Einstein-Rosen bridge 
around the origin $\mathcal O$($\bm x=0$), 
therefore the origin corresponds to the asymptotic infinity. 
The origin $\mathcal O$ is often called the ``puncture". 
Since the infinity is not a region of the spacetime, 
our initial data $\mathcal D-\{\mathcal O\}$ is homeomorphic to ${\bf T}^3$ 
with one point removed. 
In the rest of this section, we describe 
how to construct the puncture initial data with $\Lambda>0$. 

\subsubsection{constant mean curvature slice in the Kottler universe}

First, we consider a constant mean curvature(CMC) slice 
in the exact Kottler solution(Schwarzschild-de Sitter) to 
understand the puncture structure with 
$\Lambda$(see Refs.~\cite{Beig:2005ef,Nakao:1990gw} for details 
of CMC slices in the Kottler solution). 
Line elements of the Kottler solution is given by 
\begin{equation}
\dd s^2=-f(r)\dd t^2+\frac{1}{f(r)}\dd r^2+r^2\dd \Omega^2, 
\end{equation}
where
\begin{equation}
f(r)=1-\frac{2M}{r}-\frac{\Lambda r^2}{3}. 
\end{equation}
Let us consider a time slice given by 
\begin{equation}
t=h(r). 
\end{equation}
The unit normal vector field to this time slice can be expressed as  
\begin{equation}
n^\mu=\frac{1}{\sqrt{f^{-1}-fh'^2}}\left[f^{-1}(\del_t)^\mu+fh'(\del_r)^\mu\right], 
\end{equation}
where $(\del_t)^\mu$ and $(\del_r)^\mu$ are 
coordinate basis vectors. 
The CMC slice condition with the mean curvature $K$ is given by 
\begin{eqnarray}
\nabla_\mu n^\mu&=&-K
\Leftrightarrow\frac{1}{r^2}\del_r(r^2n^r)=-K\cr
&\Leftarrow& n^r=-\frac{1}{3}Kr\cr
&\Leftrightarrow& f^{-1}(1-f^2h'^2)=1/F(r;M, \Lambda, K)
=\left(1-\frac{2M}{r}-\frac{1}{3}\Lambda r^2+\frac{1}{9}K^2r^2\right)^{-1}, 
\end{eqnarray}
where we have dropped the integration constant and 
defined $F(r;M, \Lambda, K)$ as
\begin{equation}
F(r;M, \Lambda, K):=\left(1-\frac{2M}{r}-\frac{1}{3}\Lambda r^2+\frac{1}{9}K^2r^2\right). 
\end{equation}

The induced metric on the time slice is given by 
\begin{equation}
\dd\ell^2=F(r;M, \Lambda, K)^{-1}\dd r^2+r^2\dd \Omega^2. 
\end{equation}
Transformation to the isotropic coordinate can be performed as follows:
\begin{eqnarray}
\dd \ell^2&=&\Psi^4(\dd R^2+R^2\dd \Omega^2), \\
R&=&C\exp\left[\pm \int^r_{r_{\rm min}} \frac{\dd r}{r\sqrt{F(r;M, \Lambda, K)}}\right], \\
\Psi&=&\sqrt{r/R},  
\end{eqnarray}
where $r_{\rm min}$ is the throat radius given by 
$F(r_{\rm min};M, \Lambda, K)=0$. 
The minus sign branch is used in the region inside the throat. 
For this branch, 
the puncture structure requires  
\begin{equation}
R=0 ~{\rm for}~r\rightarrow \infty. 
\end{equation}
This requirement can be satisfied by setting 
\begin{equation}
K^2=3\Lambda. 
\end{equation}
This implies that we need to impose this condition near the origin
of the numerical box. 
Under this condition, the conformal factor $\Psi$ is given by 
\begin{equation}
\Psi=1+\frac{M}{2R}, 
\end{equation}
where we have set the integration constant $C$ as $C=M/2$ 
to fix the mass of the black hole measured in the infinity 
inside the black hole as $M$(see Ref.~\cite{Yoo:2012jz}).

\subsubsection{Form of $K$ and the asymptotic solution near the origin}
In this paper, we adopt the following form of $K$ 
\begin{equation}
K(\bm x)=K_{\rm \Lambda}+\left(K_{\rm b}-K_{\rm \Lambda}\right)W(R), 
\end{equation}
where $R:=|\bm x|$, $K_{\rm \Lambda}=-\sqrt{3\Lambda}$ and 
\begin{equation}
W(R)=
\left\{
\begin{array}{ll}
0&{\rm for}~0\leq R \leq \ell \\
\sigma^{-36}[(R-\sigma-\ell)^6-\sigma^6]^6&{\rm for}~\ell\leq R \leq \ell 
+ \sigma\\
1&{\rm for}~\ell+\sigma \leq R \\
\end{array}\right..
\end{equation}
$K_{\rm b}$ is a constant determined by a integrability condition discussed below. 
We set $\ell=0.1M$ and $\sigma=L-0.2M$. 
Asymptotic solution near the center is given by 
\begin{eqnarray}
&&X^i\simeq 0, \\
&&\Psi\simeq1+\frac{M}{2R}, \label{Psi-def}
\end{eqnarray}
To extract the $1/R$ divergence, we define a new variable $\psi$ as follows:
\begin{equation}
\psi(\bm x):=\Psi(\bm x)-\frac{M}{2R}\left[1-W(R)\right].  \label{psi-def}
\end{equation}
Then, the Hamiltonian constraint becomes 
\begin{equation}
\triangle\psi=\triangle \left(\frac{M}{2R}W(R)\right)
-\frac{1}{8}(\tilde L X)_{ij}
(\tilde L X)^{ij}\Psi^{-7}+\frac{1}{12}K^2\Psi^5-\frac{1}{4}\Lambda\Psi^5.  
\end{equation}

\subsubsection{Integrability condition and equations}

Integrating Eq.\eqref{eq:hamicon3}   
over the physical domain ${\cal D}-\{O\}$, 
we obtain the following equation: 
\begin{equation}
2\pi M
+\frac{1}{8}\int_{{\cal D}-\{O\}}  (\tilde L X)_{ij}(\tilde L X)^{ij}\Psi^{-7}\dd x^3
-\frac{1}{12}\left(V_1K_{\rm b}^2+2V_2K_{\Lambda}K_{\rm b}-V_3K_{\Lambda}^2\right)=0,
\label{eq:effHub1}
\end{equation}
where 
\begin{eqnarray}
V_1&:=&\int_{{\cal D}-\{O\}}W^2\Psi^5\dd^3x, \label{V1-def}\\
V_2&:=&\int_{{\cal D}-\{O\}}(1-W)W\Psi^5\dd^3x, \label{V2-def}\\
V_3&:=&V_1+2V_2. \label{V3-def}
\end{eqnarray}
This equation is the integrability condition and we choose the value of 
$K_{\rm b}$ so that Eq.~\eqref{eq:effHub1} is satisfied. That is, 
$K_{\rm b}$ cannot be freely chosen but it must be appropriately fixed through 
the numerical iteration. 

Introducing $Z$ defined by 
\begin{equation}
Z:=\del_iX^i, 
\end{equation}
we can derive the following coupled elliptic equations
\begin{eqnarray*}
\triangle\psi&=&\triangle \left(\frac{M}{2R}W(R)\right)
-\frac{1}{8}(\tilde L X)_{ij}
(\tilde L X)^{ij}\Psi^{-7}+\frac{1}{12}K^2\Psi^5-\frac{1}{4}\Lambda\Psi^5, \\
\triangle Z&=&\frac{1}{2}\del_i(\Psi^6\del^iK), \\
\triangle X^i&=&-\frac{1}{3}\del^iZ+\frac{2}{3}\Psi^6\del^iK. 
\end{eqnarray*}
We solve these equations by using the same procedure 
described in Ref.~\cite{Yoo:2012jz}. 

%%%%%%%%%%%%%%%%%%%%%%%%%%%%%%%%%%%%%%%%%%%%%%%%%%%%%%%%%%%%%%%%
\section{Marginal surfaces}
\label{sec:marginal}
%%%%%%%%%%%%%%%%%%%%%%%%%%%%%%%%%%%%%%%%%%%%%%%%%%%%%%%%%%%%%%%%

As is explicitly shown below, in our initial data, 
there are 4 marginal surfaces at most: two cosmological horizons(CHs) and 
two white hole horizons(WHs) or black hole horizons(BHs), 
if the cosmological constant is smaller 
than the Nariai bound $\Lambda =1/(9M^2)$\cite{1950SRToh..34..160N,1999GReGr..31..963N}. 
We focus on $\Lambda <1/(9M^2)$ cases in this paper. 

Hereafter, we use the words ``inner", ``outer", ``ingoing" and ``outgoing" 
based on 
the value of the numerical coordinate $\bm x$. 
That is, the innermost region is near the puncture and 
outermost region is near the boundary of the numerical box. 

The expansions of the future directed null vector fields normal to 
a 2-surface are given by 
\begin{equation}
\chi_\pm=(\gamma^{ij}-s^is^j)(\pm D_is_j-K_{ij}), 
\end{equation}
where $s^i$ is the outgoing unit vector on the initial hyper surface
which is normal to the 2-surface. 
The subscript ``$+$" means outgoing and ``$-$" means ingoing null expansion. 
If the initial hypersurface is passing through a black hole region as in the case 
$\Lambda=0$~\cite{Yoo:2012jz}, 
there are two black hole horizons(future outer trapping horizons 
in terms of Ref.~\cite{Hayward:1993wb}). 
In this case, 
the outer black hole horizon(OBH) satisfies $\chi_+=0$, and 
the inner black hole horizon(IBH) satisfies $\chi_-=0$. 
If the initial hypersurface is passing through a white hole region, 
two white hole horizons(past outer trapping) exist. 
In this case, 
the outer white hole horizon(OWH) satisfies $\chi_-=0$, and 
the inner white hole horizon(IWH) satisfies $\chi_+=0$. 
In addition, we have cosmological horizons(past inner trapping). 
The inner cosmological horizon(ICH) satisfying $\chi_+=0$ always exists inside IBH or IWH, although we may not always find it due 
to low resolution of numerical grids. 
If the box size $L$ is sufficiently large and $1/(9M^2)>\Lambda>0$, 
we can find the outer 
cosmological horizon(OCH) satisfying $\chi_-=0$ 
outside OBH or OWH. 

Equations for marginal surfaces can be rewritten as  
\begin{eqnarray}
\chi_+=0\Leftrightarrow D_is^i- K+ K_{ij}s^is^j=0 {\rm ~~for~~IWH(OBH)~and~ICH}, 
\label{eqmarg1-I}\\
\chi_-=0\Leftrightarrow 
D_is^i+ K- K_{ij}s^is^j=0 {\rm ~~for~~OWH(IBH)~and~OCH}. 
\label{eqmarg1-O}
\end{eqnarray}
Assuming that the marginal surfaces are expressed by 
$R=h(\vartheta,\varphi)$ in the spherical coordinate, 
we can rewrite Eqs.~\eqref{eqmarg1-I} and \eqref{eqmarg1-O} as 
\begin{equation}
\frac{\del^2h}{\del\vartheta^2}+\cot\vartheta\frac{\del h}{\del\vartheta}+\frac{1}{\sin^2\vartheta}
\frac{\del^2h}{\del\varphi^2}-\left(2-\eta\right)h=\eta h+S_\pm(h), 
\label{eqmarg2}
\end{equation}
where $\eta$ is a constant and $S_\pm$ is a complicated function of $h$ and geometric quantities(see e.g. \cite{Shibata:1997nc}). 
We set $\eta=3$ for CHs and $\eta=1$ for WHs(BHs). 
Although the reason is not clear, 
our experience shows that if we set $\eta=1$($\eta=3$), 
we cannot find CHs(BHs and WHs) irrespective of the initial trial for the iteration\cite{KNpriv}.

\begin{figure}[htbp]
\begin{center}
\includegraphics[scale=0.7]{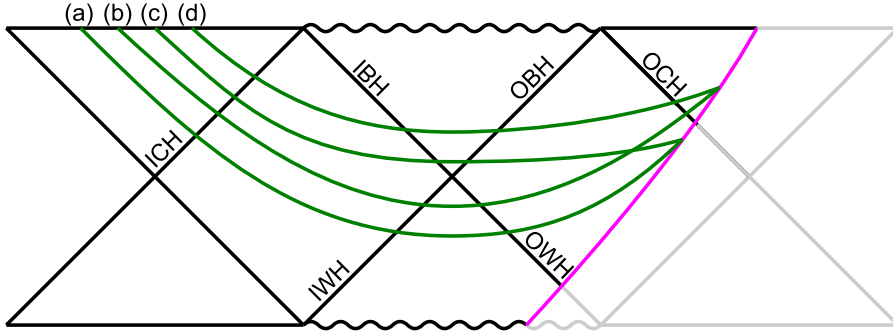}
\caption{Possible hyper-surface configurations. 
}
\label{fig:slices}
\end{center}
\end{figure}

In our settings, we may find 4 kinds of possible 
horizon configuration. 
For each case, existing horizons can be 
listed from inside to outside as follows:
\begin{itemize}
\item[(a)]
ICH, IWH, OWH
\item[(b)]
ICH, IWH, OWH, OCH
\item[(c)]
ICH, IBH, OBH
\item[(d)]
ICH, IBH, OBH, OCH
\end{itemize}
To understand these configurations, it is convenient to consider  
the Carter-Penrose diagram of the Kottler solution with an outside portion removed. 
A schematic figure of possible configurations 
is described in Fig.~\ref{fig:slices}. 
%%%%%%%%%%%%%%%%%%%%%%%%%%%<<start figure>>%%%%%%%%%%%%%%%%%%%%%%%%%%
%%%%%%%%%%%%%%%%%%%%%%%%%%%%<<end figure>>%%%%%%%%%%%%%%%%%%%%%%%%%%%

\begin{figure}[htbp]
\begin{center}
\includegraphics[scale=1]{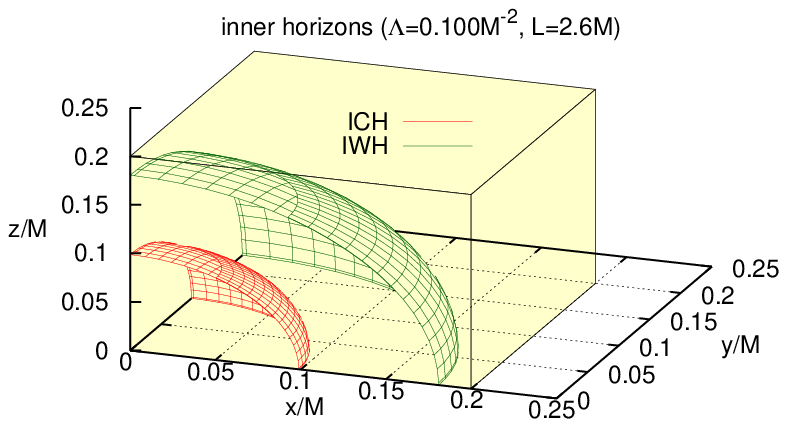}
%\hspace{1cm}
\includegraphics[scale=1]{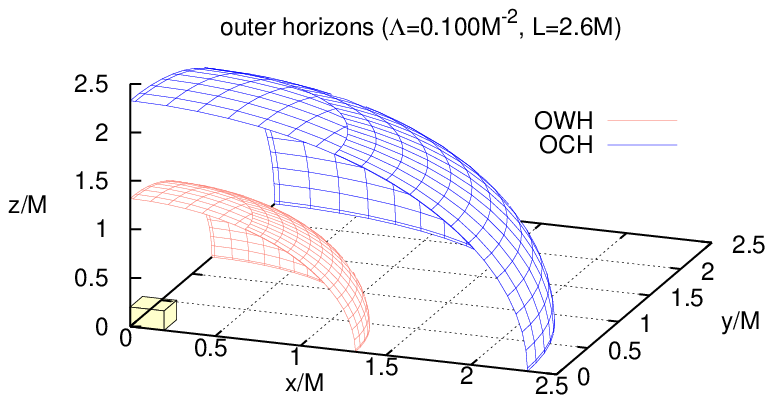}
\caption{Inner(left) and outer(right) marginal surfaces 
with $\Lambda=0.1/M^2$, $L=2.6M$. 
The left panel is a closeup figure of the central part of the right panel. 
This configuration is classified in the case (b). 
}
\label{fig_100_260}
\end{center}
\end{figure}

\begin{figure}[htbp]
\begin{center}
\includegraphics[scale=1]{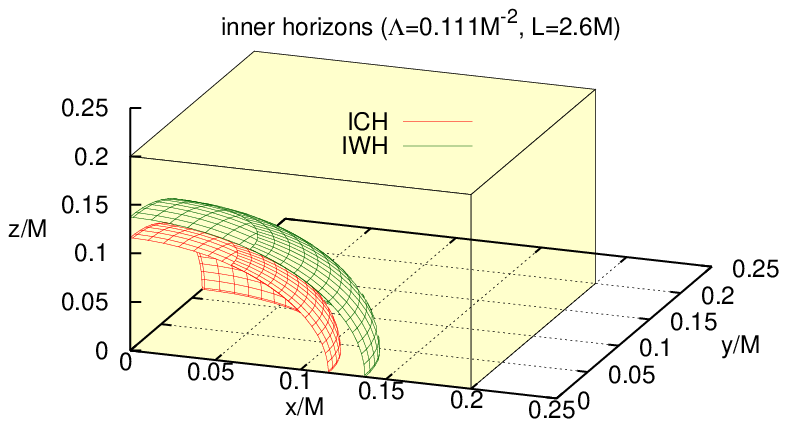}
%\hspace{1cm}
\includegraphics[scale=1]{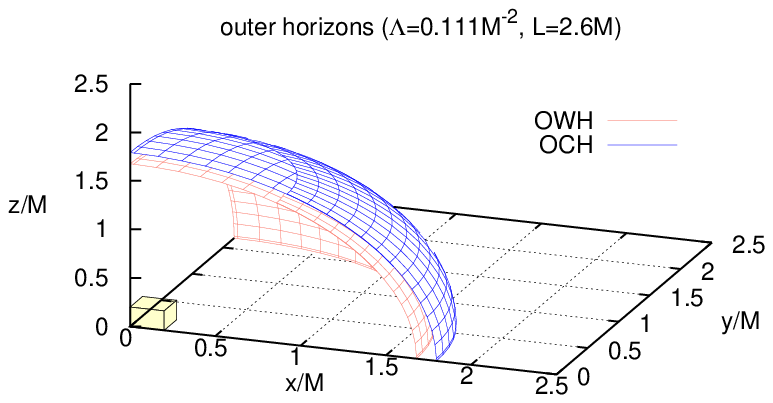}
\caption{Inner(left) and outer(right) marginal surfaces with 
$\Lambda=0.111/M^2$, $L=2.6M$.
The left panel is a closeup figure of the central part of the right panel. 
This configuration is classified in the case (b). 
}
\label{fig_111_260}
\end{center}
\end{figure}

\begin{figure}[htbp]
\begin{center}
\includegraphics[scale=1]{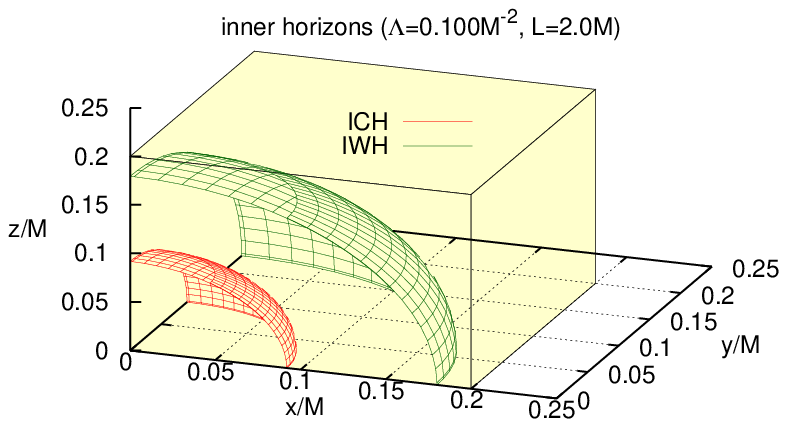}
%\hspace{1cm}
\includegraphics[scale=1]{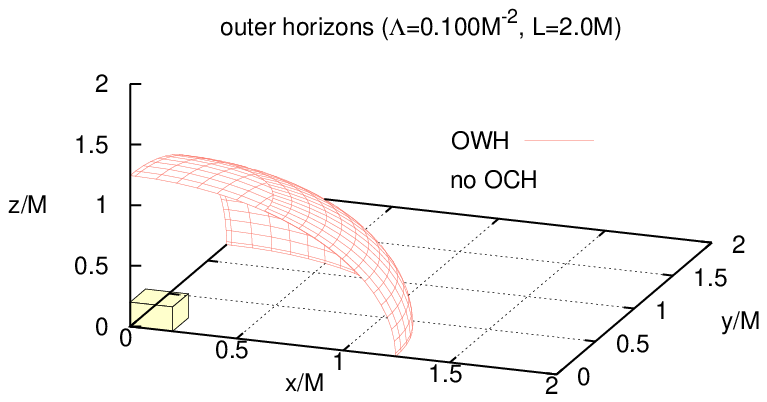}
\caption{Inner(left) and outer(right) marginal surfaces with 
$\Lambda=0.1/M^2$, $L=2M$. 
The left panel is a closeup figure of the central part of the right panel. 
This configuration is classified in the case (a). 
}
\label{fig_100_200}
\end{center}
\end{figure}

We briefly show the configuration of 
marginal surfaces on initial hypersurfaces. 
We note that, differently from the $\Lambda=0$ case in Ref.~\cite{Yoo:2012jz}, 
the hypersurface may pass through the white hole region 
for a sufficiently large value of $\Lambda$. 
In this subsection, we only show  
the cases in which the hypersurface is passing through the white hole region, that is, 
the case (a) or (b). 
As is shown in Fig.~\ref{fig_100_260}, 
there are 4 marginal surfaces for $\Lambda=0.1/M^2$ and $L=2.6M$. 
If we increase the value of $\Lambda$ to $0.111/M^2$, 
two pairs of WHs and CHs get closer as shown in Fig.~\ref{fig_111_260} 
and all marginal surfaces disappear for $\Lambda>1/(9M^2)$. 
If we decrease the value of $L$ to $2M$, OCH disappears as is 
shown in Fig.~\ref{fig_100_200}
%%%%%%%%%%%%%%%%%%%%%%%%%%%<<start figure>>%%%%%%%%%%%%%%%%%%%%%%%%%%
%%%%%%%%%%%%%%%%%%%%%%%%%%%%<<end figure>>%%%%%%%%%%%%%%%%%%%%%%%%%%%
%%%%%%%%%%%%%%%%%%%%%%%%%%%<<start figure>>%%%%%%%%%%%%%%%%%%%%%%%%%%
%%%%%%%%%%%%%%%%%%%%%%%%%%%%<<end figure>>%%%%%%%%%%%%%%%%%%%%%%%%%%%
%%%%%%%%%%%%%%%%%%%%%%%%%%%<<start figure>>%%%%%%%%%%%%%%%%%%%%%%%%%%
%%%%%%%%%%%%%%%%%%%%%%%%%%%%<<end figure>>%%%%%%%%%%%%%%%%%%%%%%%%%%%

%%%%%%%%%%%%%%%%%%%%%%%%%%%%%%%%%%%%%%%%%%%%%%%%%%%%%%%%%%%%%%%%
\section{Time Evolution}
\label{sec:evolve}
%%%%%%%%%%%%%%%%%%%%%%%%%%%%%%%%%%%%%%%%%%%%%%%%%%%%%%%%%%%%%%%%

\subsection{Settings and constraint violation}

We solve the following evolution equations by using the BSSN formalism\cite{Shibata:1995we,Baumgarte:1998te}: 
\begin{eqnarray}
\frac{\del\gamma_{ij}}{\del t}&=&-2K_{ij}, \\
\frac{\del K_{ij}}{\del t}&=&R_{ij}+KK_{ij}-2K_{ik}K^k_{~j}-\Lambda \gamma_{ij}. 
\end{eqnarray}
We describe line elements of the spacetime as follows:
\begin{equation}
\dd s^2=-N^2\dd t^2+\gamma_{ij}\left(\dd x^i+\beta^i\dd t\right)
\left(\dd x^j+\beta^j\dd t\right). 
\end{equation}
Following the previous paper\cite{Yoo:2013yea}, we use the following gauge conditions: 
\begin{eqnarray}
\left(\frac{\del}{\del t}-\beta^i\frac{\del}{\del x^i}\right)N
&=&-2N\left(K-K_{\rm c}\right), \\
\frac{\del\beta^i}{\del t}&=&B^i,\\
\frac{\del B^i}{\del t}&=&\frac{\del\tilde \Gamma^i}{\del t}-\frac{3}{4M}B^i. 
\end{eqnarray}
where $\tilde \Gamma^i:=-\del_j\tilde \gamma^{ij}$ 
and $K_{\rm c}$ is the value of $K$ at the vertex of the box. 

\begin{figure}[htbp]
\begin{center}
\includegraphics[scale=1.5]{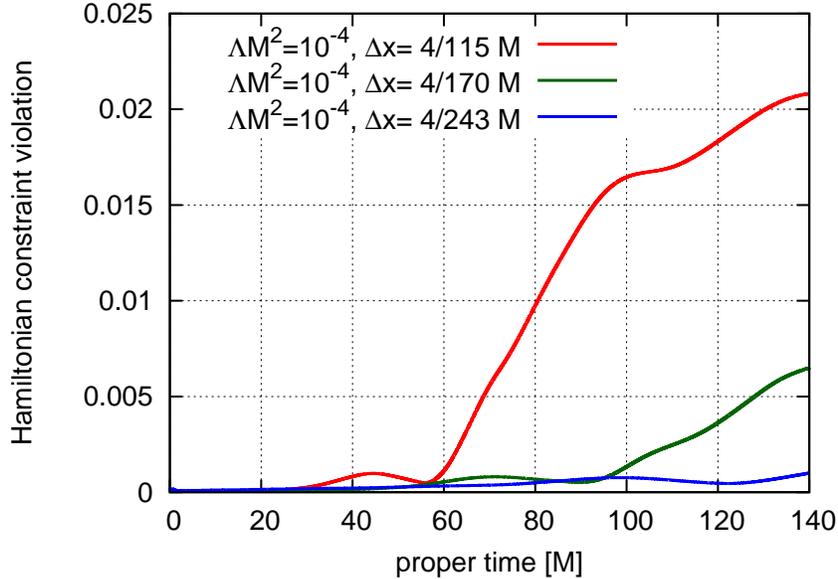}
\caption{Time evolution of the L1 norm of the Hamiltonian constraint violation. 
The value is appropriately normalized so that the minimum and maximum values 
are zero and one, respectively. 
}
\label{fig:hamcon}
\end{center}
\end{figure}

Numerical simulations are performed with 
the coordinate grid intervals $\Delta x/M$=4/51, 4/115, 4/179 and 4/243. 
The convergence of the Hamiltonian constraint violation is 
demonstrated in Fig.~\ref{fig:hamcon} for the case $\Lambda =10^{-4}/M^2$. 
We also show the convergence of the expansion law 
defining an effective scale factor in Sec.~\ref{sec:exp}. 
%%%%%%%%%%%%%%%%%%%%%%%%%%%<<start figure>>%%%%%%%%%%%%%%%%%%%%%%%%%%
%%%%%%%%%%%%%%%%%%%%%%%%%%%%<<end figure>>%%%%%%%%%%%%%%%%%%%%%%%%%%%

\subsection{Evolution of marginal surfaces}

\subsubsection{Appearance of OCH}

\begin{figure}[htbp]
\begin{center}
\includegraphics[scale=1]{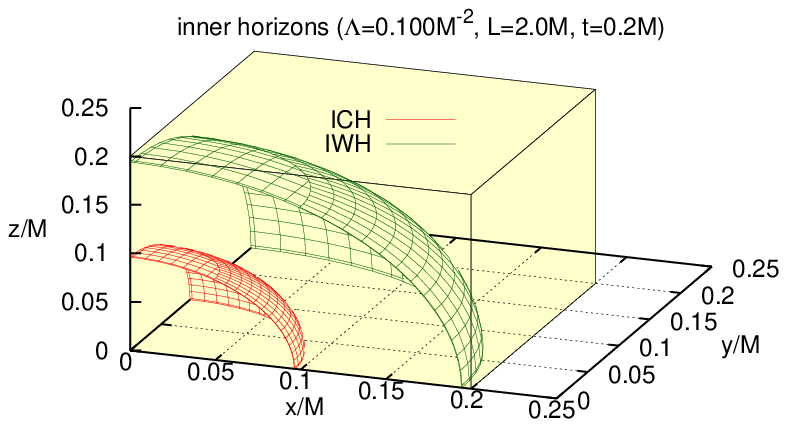}
%\hspace{1cm}
\includegraphics[scale=1]{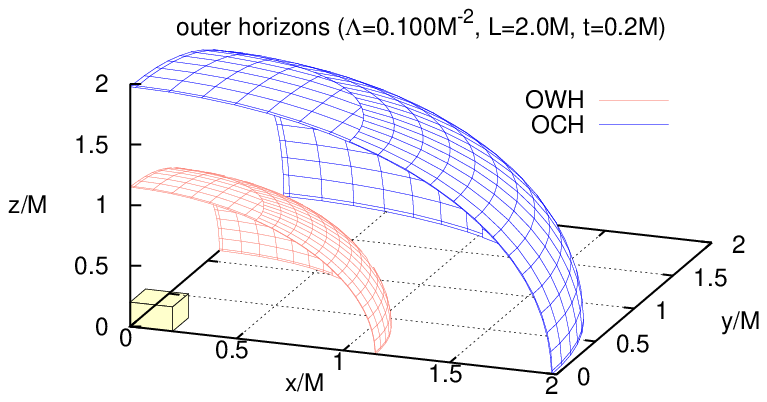}
\caption{Inner(left) and outer(right) marginal surfaces with $\Lambda=0.1/M^2$, $L=2M$ 
on the time slice given by $t\sim0.2M$.
The left panel is a closeup figure of the central part of the right panel. 
This configuration is classified in the case (b). 
}
\label{fig_100_200ev}
\end{center}
\end{figure}

One typical example($\Lambda=0.1/M^2$ and $L=2M$) is shown in Fig.~\ref{fig_100_200ev}. As is shown in Fig.~\ref{fig_100_200} there is no OCH for this case at the initial time. 
After the time evolution, at a time $t\sim0.2M$, 
the OCH appears near the boundary of the box. 
That is, the horizon configuration can change 
from the case (a) to the case (b) through time evolution(see. Fig.~\ref{fig:slices}). 
%
%%%%%%%%%%%%%%%%%%%%%%%%%%%<<start figure>>%%%%%%%%%%%%%%%%%%%%%%%%%%
%%%%%%%%%%%%%%%%%%%%%%%%%%%%<<end figure>>%%%%%%%%%%%%%%%%%%%%%%%%%%%

\subsubsection{Bifurcation surface crossing}

\begin{figure}[htbp]
\begin{center}
\includegraphics[scale=1.05]{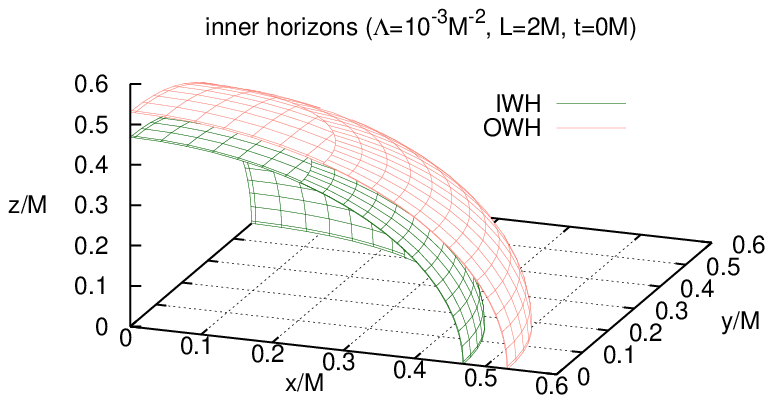}
%\hspace{1cm}
\includegraphics[scale=1.05]{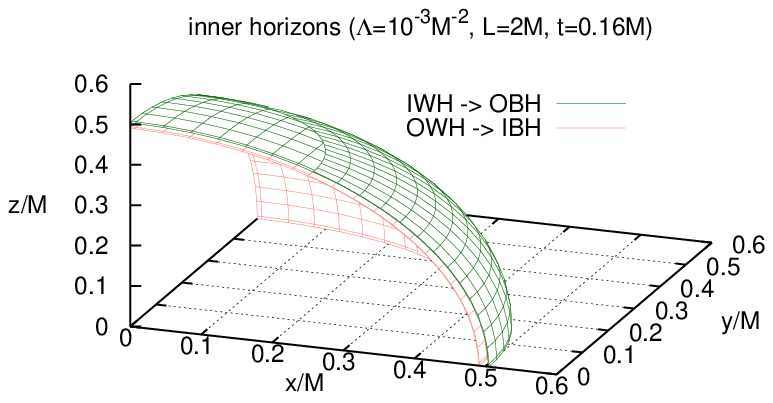}
\caption{For $\Lambda =10^{-3}/M^2$ and $L=2M$, the initial hypersurface 
is passing through the white hole region 
and these marginal surfaces are WHs(left). 
After the time evolution, 
the time slice crosses the bifurcation 2 surface and 
the marginal surfaces become BHs(right). 
Note that, while IWH and OBH satisfy $\chi_+=0$, 
OWH and IBH satisfy $\chi_-=0$. 
}
\label{fig_bifurcate}
\end{center}
\end{figure}

The other typical example is the bifurcation surface crossing. 
As is shown in Fig.~\ref{fig_bifurcate}, 
for $\Lambda =10^{-3}/M^2$ and $L=2M$, 
IWH($\chi_+=0$) exists outside OWH($\chi_-=0$). 
After the time evolution, at the time $t\sim0.15M$, 
the surface satisfying $\chi_+=0$ comes out outside the surface 
satisfying $\chi_-=0$. This implies that the 
hypersurface is passing through the black hole region and 
those surfaces are BHs. 
That is, the transitions from (a) to (c) or from (b) to (d) may happen 
through time evolution. 
%%%%%%%%%%%%%%%%%%%%%%%%%%%<<start figure>>%%%%%%%%%%%%%%%%%%%%%%%%%%
%%%%%%%%%%%%%%%%%%%%%%%%%%%%<<end figure>>%%%%%%%%%%%%%%%%%%%%%%%%%%%

\subsection{Cosmic expansion}
\label{sec:exp}

We obtain the geodesic slices parametrized by the proper time $\tau$ 
by using the same procedure described in Ref.~\cite{Yoo:2013yea}. 
Then, we calculate the effective scale factor defined by 
\begin{equation}
a_{\mathcal A}:=\sqrt{\mathcal A(\tau)}, 
\end{equation}
where $\mathcal A$ is the proper area of a surface on the geodesic slice. 
On the other hand, 
the scale factor for flat dust FLRW with $\Lambda$ 
can be written as 
\begin{equation}
a_{\rm FLRW}=a_{\rm f}\left[\frac{\left(1-\exp\left[\sqrt{3\Lambda}(t+t_{\rm f})\right]
\right)^2}%
{\left(1+\exp\left[\sqrt{3\Lambda}(t+t_{\rm f})\right]\right)^2
-\left(1-\exp\left[\sqrt{3\Lambda}(t+t_{\rm f})\right]\right)^2}\right]^{1/3}, 
\end{equation}
where we have two free parameters $t_{\rm f}$ and $a_{\rm f}$. 
We fix these parameters by fitting this form to results of numerical calculation.

\begin{figure}[htbp]
\begin{center}
\includegraphics[scale=1.5]{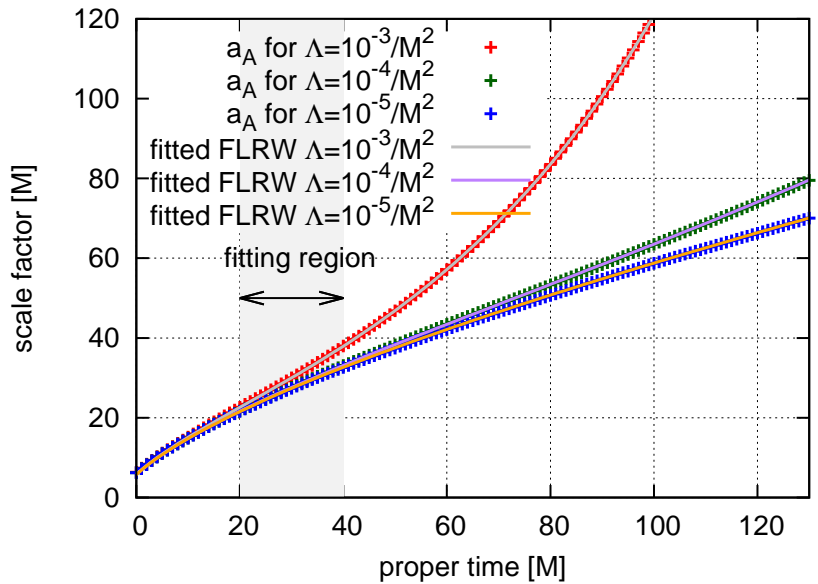}
\caption{Effective scale factors and $a_{\rm FLRW}$ for $\Lambda=10^{-3}/M^2$, 
$10^{-4}/M^2$ and $10^{-5}/M^2$, where we set $\Delta x=4/179$.  
}
\label{fig:prop}
\end{center}
\end{figure}

\begin{figure}[htbp]
\begin{center}
\includegraphics[scale=0.75]{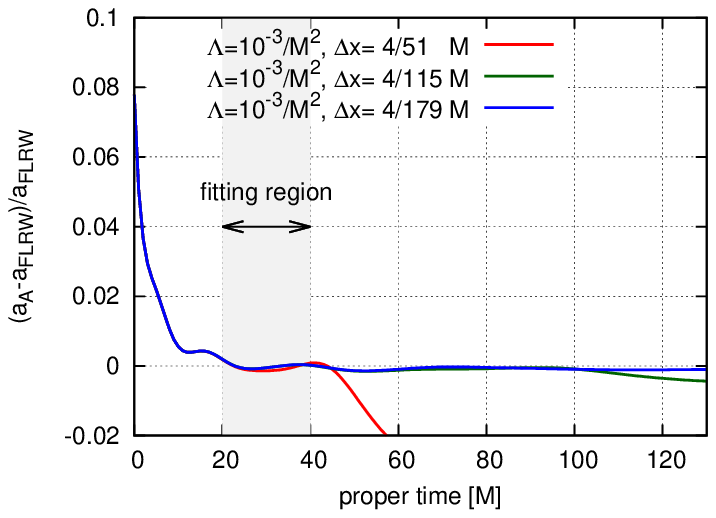}
\includegraphics[scale=0.75]{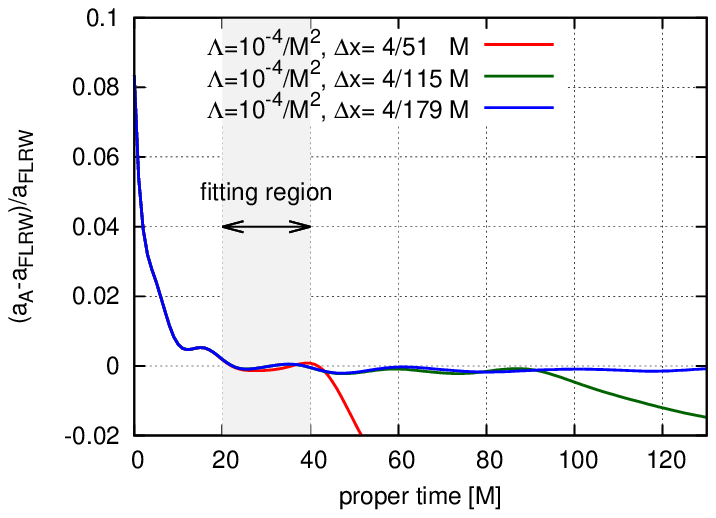}
\includegraphics[scale=0.75]{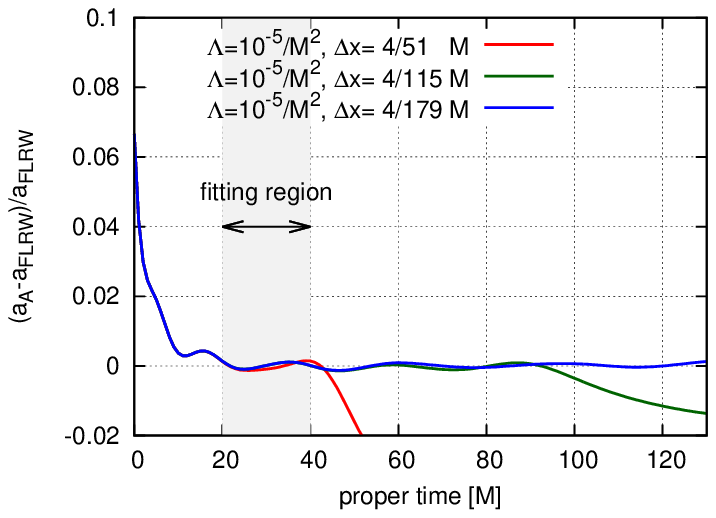}
\caption{Deviation of the effective scale factors 
from $a_{\rm FLRW}$. 
We also show the convergence of the result with smaller grid intervals in this figure. 
}
\label{fig:dev}
\end{center}
\end{figure}

Results are shown in Figs.~\ref{fig:prop} and \ref{fig:dev}. 
The evolution of the effective scale factor is well fitted by
$a_{\rm FLRW}$. 
We also show the convergence of the result with smaller grid intervals in Figs.~\ref{fig:dev}. 
The fitted values for $t_{\rm f}$ and $a_{\rm f}$ are listed in Table~\ref{tab:fit}. 
%%%%%%%%%%%%%%%%%%%%%%%%%%%<<start table>>%%%%%%%%%%%%%%%%%%%%%%%%%%
\begin{table}[htbp]
\caption{The fitted values for $t_{\rm f}$ and $a_{\rm f}$. 
}
\label{tab:fit}
\begin{tabular}{|c||c|c|c|}
\hline
$\Lambda M^2$&$10^{-3}$&$10^{-4}$&$10^{-5}$\\
\hline
\hline
$t_{\rm f}$&3.29M&3.21M&3.18M\\
\hline
$a_{\rm f}$&29.2M&62.8M&135.3M\\
\hline
\end{tabular}
\end{table}
%%%%%%%%%%%%%%%%%%%%%%%%%%%<<end table>>%%%%%%%%%%%%%%%%%%%%%%%%%%
%
This result explicitly shows that, if
the box size of the black hole universe is 
sufficiently larger than the horizon radius, 
the global expansion law can be 
well approximated by corresponding flat dust FLRW universe 
irrespective of the value of the positive cosmological constant. 
%%%%%%%%%%%%%%%%%%%%%%%%%%%<<start figure>>%%%%%%%%%%%%%%%%%%%%%%%%%%
%%%%%%%%%%%%%%%%%%%%%%%%%%%%<<end figure>>%%%%%%%%%%%%%%%%%%%%%%%%%%%
%
%%%%%%%%%%%%%%%%%%%%%%%%%%%<<start figure>>%%%%%%%%%%%%%%%%%%%%%%%%%%
%%%%%%%%%%%%%%%%%%%%%%%%%%%%<<end figure>>%%%%%%%%%%%%%%%%%%%%%%%%%%%

%%%%%%%%%%%%%%%%%%%%%%%%%%%%%%%%%%%%%%%%%%%%%%%%%%%%%%%%%%%%%%%%
\section{Summary}
\label{sec:summary}
%%%%%%%%%%%%%%%%%%%%%%%%%%%%%%%%%%%%%%%%%%%%%%%%%%%%%%%%%%%%%%%%
In this work, a black hole lattice universe model with positive 
cosmological constant has been simulated. 
The construction of puncture initial data with a positive cosmological constant 
has been described in Sec.~\ref{sec:initial}. 
The vacuum Einstein equations in a cubic box with a 
black hole in the center have been 
numerically solved with periodic boundary conditions
by using the BSSN formalism\cite{Shibata:1995we,Baumgarte:1998te}. 
Configurations of marginal surfaces on the initial hypersurfaces 
and those time evolution have been analyzed. 
We found two impressive transitions of the configuration 
in time evolution: appearance of the outer cosmological horizon and 
the bifurcation surface crossing. 
Finally, comparing the effective scale factor defined by the surface area 
and the scale factor for the corresponding flat dust FLRW universe, 
we have concluded that the expansion law of 
the black hole universe can be well approximated by 
that of the  corresponding flat dust FLRW universe 
in the sufficiently late time 
irrespective of the value of the cosmological constant. 

\section*{Acknowledgements}
We thank T. Tanaka, M. Sasaki and K. Nakao for 
helpful discussions and comments.

%\bibliographystyle{h-physrev5}
%\bibliographystyle{h-physrev5-title}
%\bibliography{../bibfiles/BHuni}

\end{document}